\documentclass[aps,prb,superscriptaddress,floatfix,twocolumn]{revtex4-1}
\usepackage{amsmath}
\usepackage{amssymb}
\usepackage{times}
\usepackage{graphicx}
\usepackage{dcolumn}
\usepackage{natmove}
\usepackage{hyperref}
\usepackage[usenames,dvipsnames]{color}
\hypersetup{
	breaklinks=true
}

\begin{document}

\title{Spectroscopy of multi-electrode tunnel barriers}

\author{A. Shirkhorshidian}
\email{ashirkh@sandia.gov}
\affiliation{University of New Mexico, Albuquerque, New Mexico 87131, USA}
\affiliation{Sandia National Laboratories, Albuquerque, New Mexico 87185, USA}

\author{John King Gamble}
\affiliation{Center for Computing Research, Sandia National Laboratories, Albuquerque, New Mexico 87185, USA}

\author{L. Maurer}
\affiliation{Center for Computing Research, Sandia National Laboratories, Albuquerque, New Mexico 87185, USA}

\author{S. M. Carr}
\affiliation{Sandia National Laboratories, Albuquerque, New Mexico 87185, USA}

\author{J. Dominguez}
\affiliation{Sandia National Laboratories, Albuquerque, New Mexico 87185, USA}

\author{G. A. Ten Eyck}
\affiliation{Sandia National Laboratories, Albuquerque, New Mexico 87185, USA}

\author{J. R. Wendt}
\affiliation{Sandia National Laboratories, Albuquerque, New Mexico 87185, USA}

\author{E. Nielsen}
\affiliation{Center for Computing Research, Sandia National Laboratories, Albuquerque, New Mexico 87185, USA}

\author{N. T. Jacobson}
\affiliation{Center for Computing Research, Sandia National Laboratories, Albuquerque, New Mexico 87185, USA}

\author{M. P. Lilly}
\affiliation{Sandia National Laboratories, Albuquerque, New Mexico 87185, USA}
\affiliation{Center for Integrated Nanotechnologies, Sandia National Laboratories, Albuquerque, New Mexico 87185, USA}

\author{M. S. Carroll}
\email{mscarro@sandia.gov}
\affiliation{Sandia National Laboratories, Albuquerque, New Mexico 87185, USA}

\begin{abstract} 
Despite their ubiquity in nanoscale electronic devices, the physics of tunnel barriers has not been developed to the extent necessary for the engineering of devices in the few-electron regime.
This problem is of urgent interest, as this is the precise regime into which current, extreme-scale electronics fall.
Here, we propose theoretically and validate experimentally a compact model for multi-electrode tunnel barriers, suitable for design-rules-based engineering of tunnel junctions in quantum devices.
We perform transport spectroscopy at $T=4$~K, extracting effective barrier heights and widths for a wide range of biases,
using an efficient Landauer-B\"uttiker tunneling model to perform the analysis.
We find that the barrier height shows several regimes of voltage dependence, either linear or approximately exponential. 
The exponential dependence approximately correlates with the formation of an electron channel below an electrode.  
Effects on transport threshold, such as metal-insulator-transition and lateral confinement are non-negligible and included. 
We compare these results to semi-classical solutions of Poisson's equation and find them to agree qualitatively.
Finally, we characterize the sensitivity of a tunnel barrier that is raised or lowered without an electrode being directly above the barrier region.
\end{abstract}

\maketitle

\section{Introduction}
Electron tunneling is a phenomenon that is generally well-understood having been studied in numerous configurations over many decades \cite{sze1981physics,davies1997}.  In recent semiconductor devices, tunneling effects are becoming increasingly important as dimensions of devices are scaled to their ultimate limits. Many Beyond-Moore's-Law device concepts, for example, invoke devices for which tunneling is a central element of the device (\emph{e.g.}, tunnel FETs) and the tunnel barrier is necessarily voltage tunable \cite{ionescu2011}.  In practice, numerous phenomenological tunnel models and interpretations of the models are invoked to design or extract barrier heights and widths (\emph{e.g.}, WKB \cite{simmons1963}, Fowler-Nordheim \cite{lenzlinger1969,huisman2009}, and Non-equilibrium Green's function \cite{ilatikhameneh2015,gao2014}).  Despite the profound foundational understanding of tunneling, the details of the barrier and its geometry are known to produce non-trivial effects \cite{friesen2013}.  This challenges both design and analysis of tunnel-based devices.

Voltage-tuned barriers represent a significant increase in complexity in contrast to barriers formed by static material barriers (\emph{e.g.}, heterostructures). In these circumstances the barriers deform according to the applied voltages and it is unclear how effective simple phenomenological models that assert, for example, two parameter descriptions like width and barrier height are. Furthermore, the functional dependence of these parameters on voltage is not well understood. Observations of linear dependence of barrier height on voltage have been conjectured for some geometries\cite{maclean2007}, but it is not clear under what conditions this holds or what to expect in general.

In this work, we use tunneling spectroscopy at 4~K to characterize the voltage dependence of a barrier in a lateral, electrostatically gated nanostructure using a MOS gate stack.  
In particular, we methodically examine and parameterize the effect of multiple electrodes on how the barrier deforms in this geometry, which extends previous studies in this area.  
The lateral, electrostatically gated MOS nanostructure is a model system that applies abstractly to many other systems while also being directly informative to the ubiquitous MOS system. 
We show that a two-parameter model (width and barrier height) is sufficient to describe the observed tunneling over a wide range of voltages as long as the barrier height and width are dependent on electrode voltages. 
We find that the functional barrier height dependence has multiple regimes, ranging from linear to approximately  exponential. 
The non-linear regime correlates with the formation of an electron channel in this MOS enhancement mode configuration. 
Changes in local bias from neighboring electrodes produce voltage shifts in the overall behavior, which otherwise can be described by similar parameterization. 
This suggests the possibility that tunnel barriers might be simulated with a relatively simple compact model, which would greatly simplify modeling of more complex laterally gated nanostructures (\emph{e.g.}, quantum dot networks \cite{zajac2015}).  
Finally, we characterize the tunnel barrier control that can be achieved using a reservoir enhancement gate instead of a dedicated barrier gate (\emph{i.e.}, a more efficient single metal layer QD layout \cite{rudolph2016}). 

This paper is organized as follows: In Sec.~\ref{sec:experiment} we describe the device fabrication and transport measurements of the tunnel barrier. In addition we discuss calculations of the electron density and Fermi energy in the leads and consider effects such as metal-insulator transition and lateral quantum confinement. Sec.~\ref{sec:theory} describes the 1D barrier model and we present two methods for computing transmission through the barrier. We then discuss the results of fitting the 1D barrier model to the data and compare these results to electrostatic simulations in Sec.~\ref{sec:results}.

\begin{figure*}[tb]
     \includegraphics[width=0.95\textwidth]{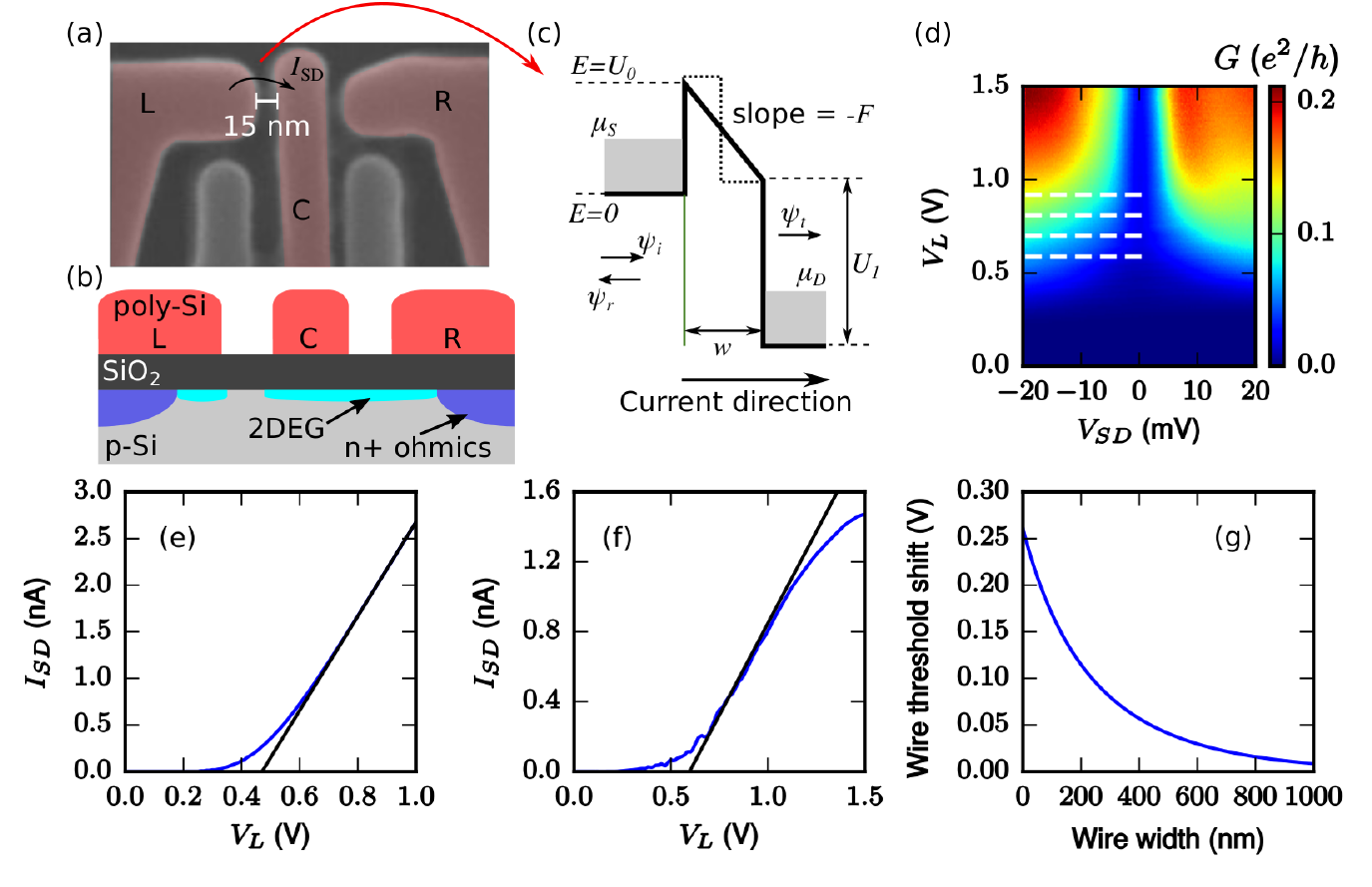} 
   \caption{ \label{fig1}Device geometry and threshold measurements.
   (a): False-color SEM of the measured device. 
   Only the red-colored gates (L, C, and R) are used; all other gates are grounded. 
   (b): Cross-sectional schematic of the device (not to scale). 
   The device stack includes a p-Si substrate (light grey), 35 nm thermally grown SiO$_{2}$ (dark grey), and 200 nm poly-Si gates (red). 
   The L and R gates spread out to overlap n$^{+}$ ohmics (purple) 
   (c): Energy diagram as described in the text showing conduction band minimum energy vs. distance along the direction of current flow. 
   (d): Differential conductance for $V_{R} = 2.13$~V and $V_{C} = 1.60$~V. 
   Dashed lines correspond to the line cuts shown in Fig.~\ref{fig3}(a). 
   (e): Measured current in the field of the device showing a field threshold of 0.48~V. 
   (f): Measured current through nanostructure for $V_{R} = 2.13$~V and $V_{C} = 5.00$~V giving a threshold of 0.60~V.   The  $V_{C}$ value is chosen because it is representative of voltages above $\sim$2.00~V.  Threshold decreases slightly for $V_{C} < 2.00$~V.
   (g): Calculated threshold shift as function of wire width for a field threshold of 0.48~V.  Lateral quantum confinement is predicted to produce a threshold shift. Details of the calculation are found in Appendix~\ref{sec:confinement}.}
\end{figure*}

\section{Experimental approach}
\label{sec:experiment}
The tunnel barriers are formed in a metal oxide semiconductor (MOS) field effect transistor (FET) inversion channel at a Si/SiO$_{2}$ interface. The starting material is a p-type Si substrate produced by a float-zone process and with a doping concentration of $\approx10^{14}$ cm$^{-3}$ boron. The structures are made by first forming heavily doped n+ ohmics using ion implantation of As. Next, a 35~nm Si$\text{O}_{2}$ gate dielectric is grown by thermal oxidation at 900$^{\circ}$C. Degenerately doped n-type poly-Si is formed over the SiO$_{2}$ to complete the MOS gate stack. The gate structures are patterned using electron beam lithography and a dry etch using a fabrication process described in more detail elsewhere \cite{singh2016}. 
A scanning electron microscope (SEM) image of the nanostructure gates and cross-sectional schematic view of the device are shown in Figs.~\ref{fig1}(a-b).
As seen in  Fig.~\ref{fig1}(a) the relevant gates are separated from each other by approximately 15~nm.  The inversion channel is formed under the poly-Si enhancement gates and the barriers are formed at gaps between the poly-Si gates.

A simplified model of the conduction band at low temperature is shown in  Fig.~\ref{fig1}(c). A high-density electron channel forms when gates L, C, and R are biased above threshold. If we decrease the bias on one of the gates below threshold, electrons are locally depleted and the conduction band edge rises above the Fermi level forming a tunnel barrier with source and drain leads to the left and right of the barrier. The extent to which the Fermi level in the leads is above the conduction band edge is proportional to the electron density. Assuming a 2D density of states (DOS), the difference between the Fermi energy $E_{f}$ and conduction band edge $E_{C}$ is
\begin{equation}
E_{f}-E_{C}=\frac{\pi\hbar^{2}n}{m_t g_{v}}, \label{Fermi}
\end{equation}
where $n$ is the 2D electron density, $g_{v}$ is the valley degeneracy, and $m_t$ is the electron transverse effective mass. Spin degeneracy, $g_{s}=2$, is already assumed. We use a valley degeneracy $g_{v}=2$ and an effective mass of $m_t=0.19m_{0}$, appropriate for 2D electron inversion layers in Si(100) \cite{beenakker1991}, where $m_{0}$ is the free-electron mass. 

We measure the differential conductance $G=\mathrm{d}I / \mathrm{d}V_{SD}$ using a lock-in amplifier with an AC excitation voltage of 100~$\mu$V at a modulation frequency of 149~Hz in combination with a DC source-drain bias $V_{SD}$. We also monitor the DC source-drain current $I_{SD}$ using a digital multi-meter. All measurements are performed at 4~K by immersing the sample in liquid He. A plot of the differential conductance as a function of $V_{L}$ and $V_{SD}$ for $V_{R}=2.13$~V and $V_{C}=1.60$~V is shown in Fig.~\ref{fig1}(d). Disorder such as charge defects in the gate oxide \cite{nordberg2009}, impurities \cite{lansbergen2008}, or strain from the gate stack \cite{thorbeck2015} can cause sub-threshold resonances in these plots. The absence of resonances suggests a single tunnel barrier between the source and drain leads. We also note that the combination of the contact resistance and sheet resistance of the channel leads to approximately 10~k$\Omega$, which limits the maximum conductance.

To calculate the tunneling current, we first need to estimate the voltage dependence of the Fermi energy in the leads.  We first estimate the electron density dependence on voltage in the field of the device. In the field we exclude effects due to the nanoscale gates, such as fringing fields and lateral quantum confinement (\emph{i.e.}, we assume a bulk poly-Si gate).  We identify a field threshold of 0.48~V, Fig.~\ref{fig1}(e).  At cryogenic temperatures there is a metal-insulator transition for which there is a critical density of electrons necessary before appreciable conduction starts. We find a critical density of $n_{crit} = 3.1\times10^{11}$~cm$^{-2}$ measured in a Hall bar with nominally similar oxide/silicon interface as used in this tunnel barrier experiment. We estimate the critical density by extrapolating the low density mobility to zero mobility \cite{beveren2010,borselli2011a}. Assuming that $n = n_{crit}$ at $V_{gate} = V_{th}$ where $V_{th}$ is the threshold voltage then the electron density as a function of gate voltage is
\begin{equation}
 n\left(V_{gate}\right) = \frac{C_{ox}}{e}(V_{gate}-V_{th}) + n_{crit}, \label{density}
\end{equation}
Where $C_{ox}$ is the oxide capacitance per unit area and $e$ is the electron charge. The electron density and Fermi energy can be calculated in the field using Eqs.~\ref{density} and~\ref{Fermi}. The accounting of the metal-insulator transition accommodates fitting over the full range of voltages including the sub-threshold region using a simple 1D capacitance model to approximate the density in the leads. We note that this compact model approximation for the electron density is not strictly accurate in the sub-threshold region. Conductances at densities below $n_{crit}$ should be negligible. However, since the transmission is relatively weakly dependent on the lead density, we don't expect significant error for extending the capacitance model for the densities below $n_{crit}$.

We also use the threshold voltage and critical density to extract an effective fixed charge density near the interface of $Q_{f} = 3.9\times10^{10}$~cm$^{-2}$, assuming a standard MOS threshold calculation adjusting parameters for 4~K \cite{muller2002device}. The details of this calculation are in Appendix~\ref{sec:threshold}. The fixed charge would notably be below zero if the critical density was not considered.  Hall measurements of mobility dependence on density can also be used to quantify a different but related charge density, the Coulomb scattering charge density.  We find that the slope of the leading edge of the mobility dependence on density fits with a scattering charge density of $7.4\times10^{10}$~cm$^{-2}$, in near quantitative agreement with the extracted fixed charge density. The scattering charge density is extracted through interpolation between mobility vs. density curves calculated for the relevant range of scattering charge densities \cite{tracy2009,das-sarma2013,stern1980}. The Hall measurements also show a peak mobility of 5950~cm$^{2}$/(V$\cdot$s) and an estimated surface roughness and correlation length of 2.2~{\AA} and 24~{\AA}, respectively, from the falling edge dependence \cite{mazzoni1999}. 

The electron density in the leads of the tunnel barrier will differ from that in the field due to two effects: fringing field and lateral quantum confinement.  
We calculated electrostatic modifications of threshold due to fringing fields numerically and were found to be a small, less than 10\% effect.
The effect of quantum confinement is more significant \cite{lu2016}. 
We calculated the confinement threshold shift for this geometry over various wire widths, see Fig.~\ref{fig1}(g) and Appendix~\ref{sec:confinement}. 
We measure a threshold of 0.60$\pm$0.05~V for the nanostructure (\emph{i.e.}, $V_{R}$ = 2.13~V, $V_{C}$ = 5.00~V, and $V_{L}$ is increased).  This threshold is the linear extrapolation to zero current, Fig.~\ref{fig1}(f).   The threshold shift falls within a range that would be expected for the $\sim$70~nm width of the tunnel barrier lead.  The electron density in the leads can, therefore, be estimated using the field threshold combined with an offset from quantum confinement.

\section{Theoretical model}
\label{sec:theory}
We now examine how well a 1D barrier model and a voltage dependent parameterization of barrier height and width fits the observed tunneling. At zero source-drain bias, we assume a rectangular barrier with a width $w$, a left barrier height $U_{0}$, and a right barrier height $U_{1}$. The two barrier heights allow for asymmetric plateaus in the leads. When a source-drain bias $V_{SD}$ is applied, the chemical potentials or quasi-Fermi levels in the source $\mu_{S}$ and drain $\mu_{D}$ separate by an amount proportional to $V_{SD}$. The potential in the barrier region then varies linearly with a slope equal to $–F$, where $F$ is the electric field due to the source-drain bias, as shown in Fig.~\ref{fig1}(c). Thus, for non-zero source-drain bias, the wave-functions in the barrier region are Airy functions. In the leads the wave-functions are propagating plane waves with an incident wave $\psi_{i}$, a reflected wave $\psi_{r}$, and a transmitted wave $\psi_{t}$.

We use the Landauer-B\"{u}ttiker formalism to model the tunneling transport. Assuming low temperature, the Fermi functions in the source and drain leads can be approximated as step functions and the current through the barrier is given by \cite{datta1997electronic}
\begin{equation}
I=\int_{\mu_{0}}^{\mu_{S}} \mathrm{d}E \, \frac{2e}{h}T(E)M(E). \label{Eq1}
\end{equation}
Here, $T(E)$ is the transmission coefficient of an electron with energy $E$, $M(E)$ is the total number of available transverse modes which depends on the DOS in the leads, and $\mu_{0}=\max \left( \mu_{D},0 \right)$. 
Thus, the product $T(E)M(E)$ gives the total transmission summed over the transverse modes. 

We examine both a numerical and approximate analytic solution to computing the transmission.  Both are sufficiently efficient to find good fits for barrier height and width for each gate voltage. The numerical approach uses the analytical, piecewise solution to Schr\"{o}dinger's equation and then numerically solves the boundary matching problem as a system of equations. This approach gives us an exceptionally fast forward solve (contrasted against fully numerical approaches) to facilitate non-linear inference over a large experimental data set. For more details, see Appendix~\ref{sec:formalism}.

\begin{figure}[tb]
   \includegraphics[width=0.95\columnwidth]{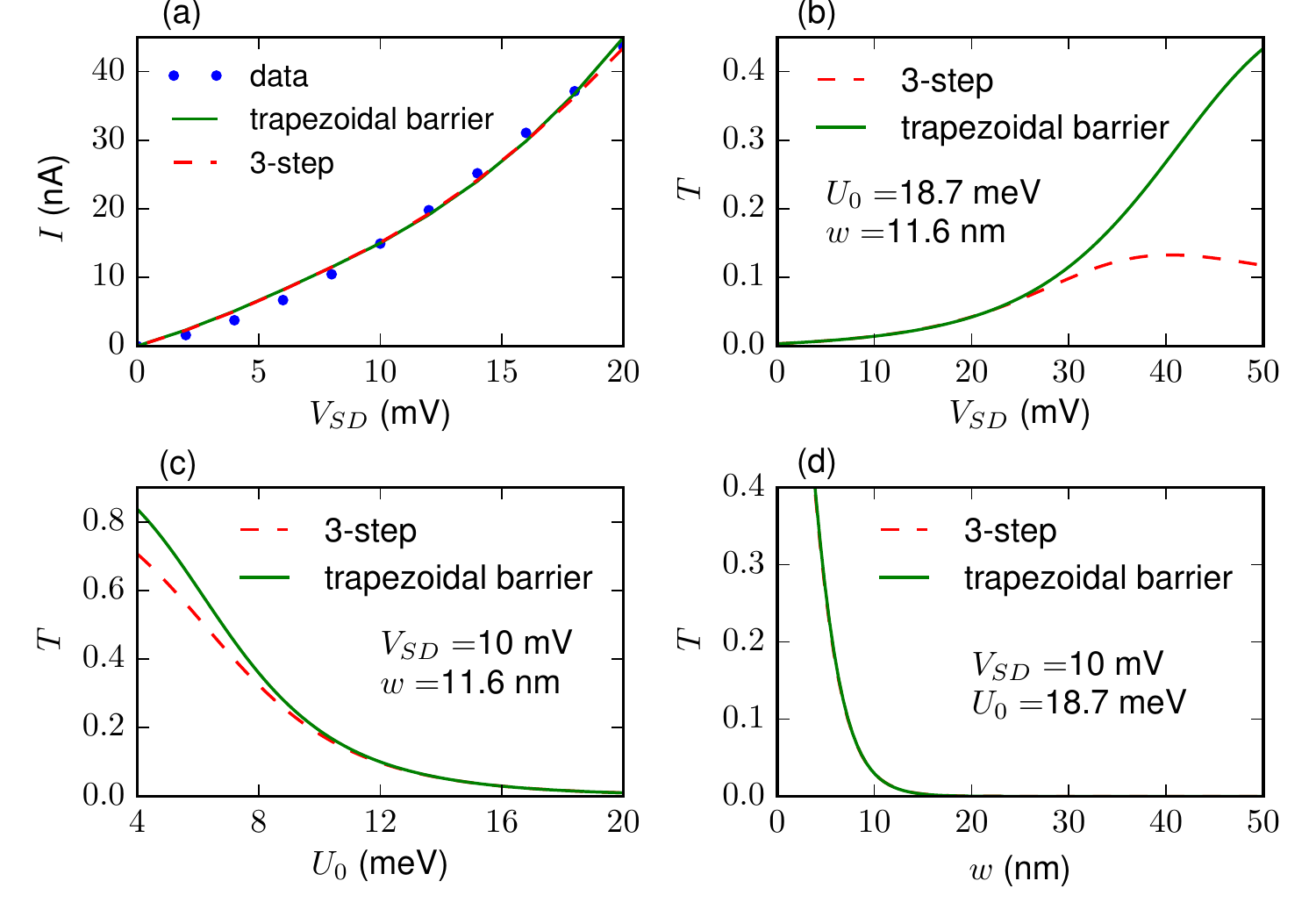} 
   \caption{\label{fig2}Comparison between the trapezoidal and three-step barrier models.
   (a): Line-cut data from Fig.~\ref{fig1}(d) for $V_{L} = 0.70$~V together with fit using the trapezoidal barrier model.
   The extracted fit parameters are used to calculate the current using the three-step model. 
   We then compare the resulting transmission coefficients $T$ as a function of (b) $V_{SD}$, (c) $U_{0}$,  and (d) $w$.}
\end{figure}

Reasonably good quantitative agreement can also be provided through solving the transmission coefficient by approximating the linearly varying potential in the barrier region with two potential steps, see Fig.~\ref{fig1}(c). One step has a fixed height of $U_{0}$ while the other has a variable step height that depends on $V_{SD}$. The transmission coefficient is then solved using transfer matrices and yields a simple, analytic formula for transmission that is a good approximation to the trapezoidal barrier problem over a wide parameter range. More details are provided in Appendix~\ref{sec:threeStep}. We find that both 1D models can match the measured current dependences on voltage, Fig.~\ref{fig2}(a).  We further find that the three-step model agrees well with the full numeric solution for lower $V_{SD}$ relative to the barrier $U_{0}$, Fig.~\ref{fig2}(b-c).  
For the rest of this paper we will use the numerical solution to the trapezoidal model for the more accurate quantitative analysis.

We note that the three-step and trapezoidal barrier models are useful alternatives to traditional approximate models of tunneling phenomena, such as the WKB approximation and Fowler-Nordheim tunneling. 
Our approach avoids unnecessary (and often unjustified) assumptions by exactly solving Schr\"{o}dinger's equation and more accurately accounting for the lead geometry.  
Over the course of a numerical fit, the assumptions of these approximations can easily become violated, invalidating the parameter extraction. 
Both our numerical solution to the trapezoidal barrier and the three-step model are simple enough to allow rapid estimates of the barrier height and width from transport measurements using a modern computer.

\section{Results and Discussion}
\label{sec:results}
\begin{figure*}[t]
   \includegraphics[width=0.95\textwidth]{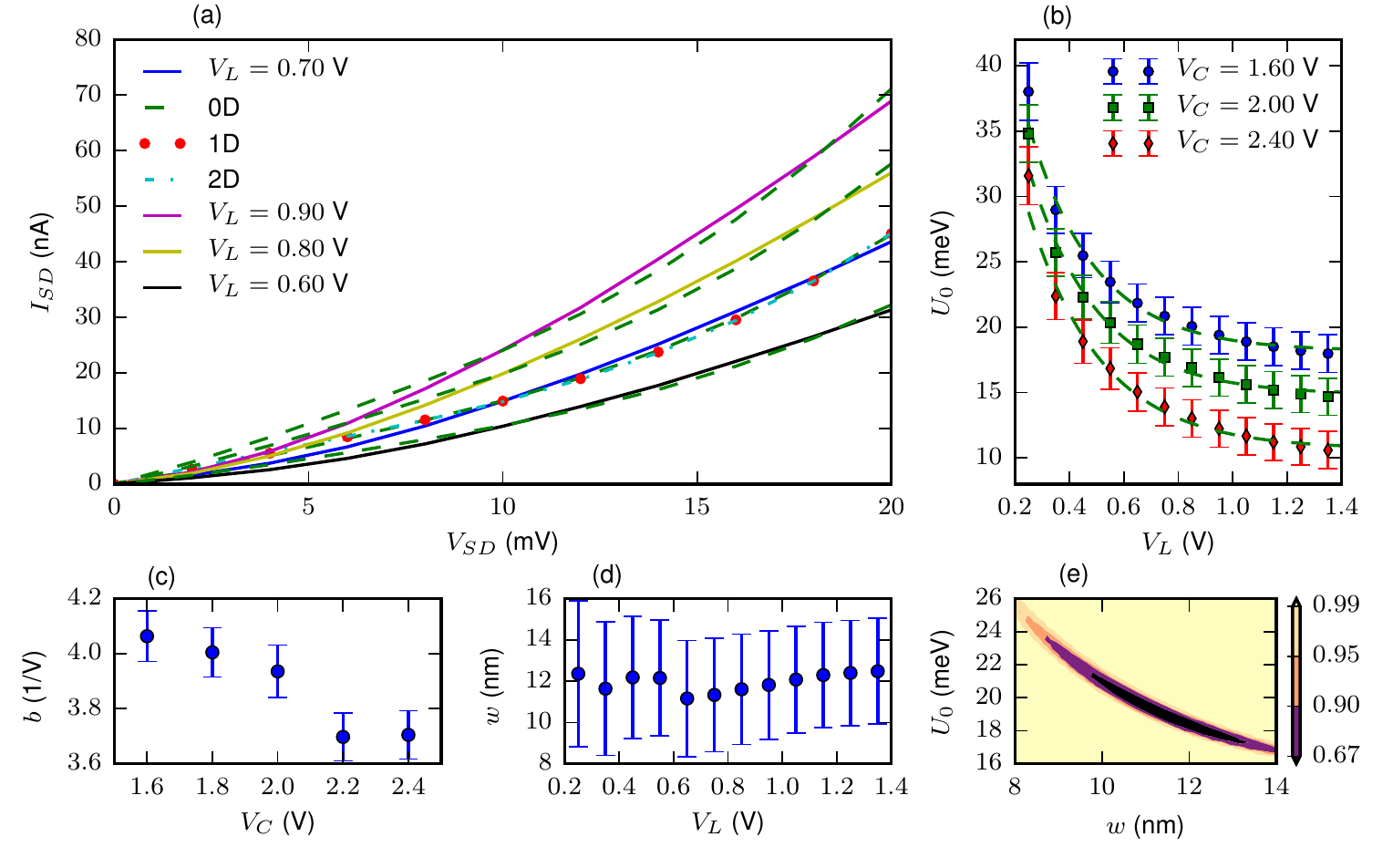} 
   \caption{\label{fig3}Parameter extraction from experimental data.
   (a): Line-cut data from Fig.~\ref{fig1}(d) together with fits using semi-analytical model. For $V_{L}=0.70$~V we compare 0D, 1D, and 2D DOS in the leads. 
   Note that the data are plotted as the absolute value of the DC source-drain current $I_{SD}$ versus the absolute value of $V_{SD}$. 
   (b): Extracted barrier height $U_{0}$ as a function of gate $V_{L}$ for a fixed width $w$ = 12~nm. Dashed lines are fits to an exponential function. Curves are artificially offset vertically to view each case independently. For $V_{C}=1.60$~V the offset is +3~meV and for $V_{C}=2.40$~V the offset is -3~meV. There is a high degree of orthogonal control of the barrier depending strongly on L and weakly on C. 
   (c): Extracted exponent from fits shown in panel (b) as a function of $V_{C}$. The average exponent is 3.9~V$^{-1}$. 
   (d): Extracted barrier width $w$ as a function of gate $V_{L}$ for a non-fixed $U_{0}$.  
   (e): Confidence region plot as function of $w$ and $U_{0}$ for $V_{L}=0.70$~V. Contours correspond to 67\%, 90\%, 95\%, and 99\% confidence levels.}
\end{figure*}

Figure~\ref{fig3}(a) shows line-cuts at $V_{L} = $ 0.90, 0.80, 0.70 and 0.60~V from the data in Fig.~\ref{fig1}(d). 
We treat the barrier height $U_{0}$ and width $w$ as free parameters and fit the trapezoidal barrier model to the data by minimizing a chi-squared test statistic, resulting in maximum likelihood estimates (MLEs) of the parameters under the assumption of independent and identically distributed, Gaussian errors. 
These fits are shown in Fig.~\ref{fig3}(a). 
The MLEs and expected errors for the parameters are shown in Table~\ref{tab:extractParams}. 

The expected errors were computed by constructing the profile likelihood function (separately, for each parameter), and numerically determining a confidence level of $95\%$.
To do this, we make the standard assumption that the difference in the profile likelihood function from the MLE value should be $\chi^2(1)$ distributed.
Unless otherwise noted, we assume throughout this paper that the experimental standard error in the observed currents is $\delta I = 0.1 \cdot \text{max}\left(I_{SD}\right)$, where $\text{max}\left(I_{SD}\right)$ is the maximum source-drain current for a given gate voltage. 
From Table~\ref{tab:extractParams}, we see that the barrier width seems to be fixed at about 12~nm while the barrier height and its uncertainty increase as $V_{L}$ decreases. 

We repeat the fitting procedure for different $V_{L}$ values. Fig.~\ref{fig3}(b) shows a plot of the extracted barrier height as a function of $V_{L}$ for a fixed width of 12~nm and three values of $V_{C}$. In this case we assume $\delta I = 0.3 \cdot \text{max}\left(I_{SD}\right)$ because the errors were found to be smaller for a one-parameter fit. We find that at high $V_{L}$ the barrier height varies linearly with gate voltage. However for low $V_{L}$ the barrier height increases non-linearly. 

Overall, we observe, that the dependence fits an exponential function of the form $f(x)=a\mathrm{e}^{-bx}+c$. 
In Fig.~\ref{fig3}(c), we compare the exponent $b$ as a function of $V_{C}$, finding that it varies between 4.1~V$^{-1}$ to 3.7~V$^{-1}$ as $V_{C}$ changes from 1.60~V to 2.40~V. The average exponent is 3.9~V$^{-1}$. This non-linear increase in barrier height approximately coincides with the regime for which charge density is changing exponentially as the electron channel is forming under the gate.

\begin{table}[tb] 
\caption{\label{tab:extractParams}Extracted barrier parameters $U_{0}$ and $w$ for fits shown in Fig.~\ref{fig3}(a).}
\begin{ruledtabular}
\begin{tabular}{cccc}
$V_{L}$ (V) & DOS of leads & $U_{0}$ (meV) & $w$ (nm)\\
\hline
0.90 & 0D & 17$\pm$3 & 12$\pm$3\\
0.80 & 0D & 18$\pm$4 & 11$\pm$3\\
0.70 & 0D & 19$\pm$5 & 12$\pm$3\\
0.70 & 1D & 27$\pm$8 & 12$\pm$3\\
0.70 & 2D & 29$\pm$9 & 12$\pm$3\\
0.60 & 0D & 20$\pm$5 & 12$\pm$3\\
\end{tabular}
\end{ruledtabular}
\end{table}

\begin{figure}[tb!]
   \includegraphics[width=0.95\columnwidth]{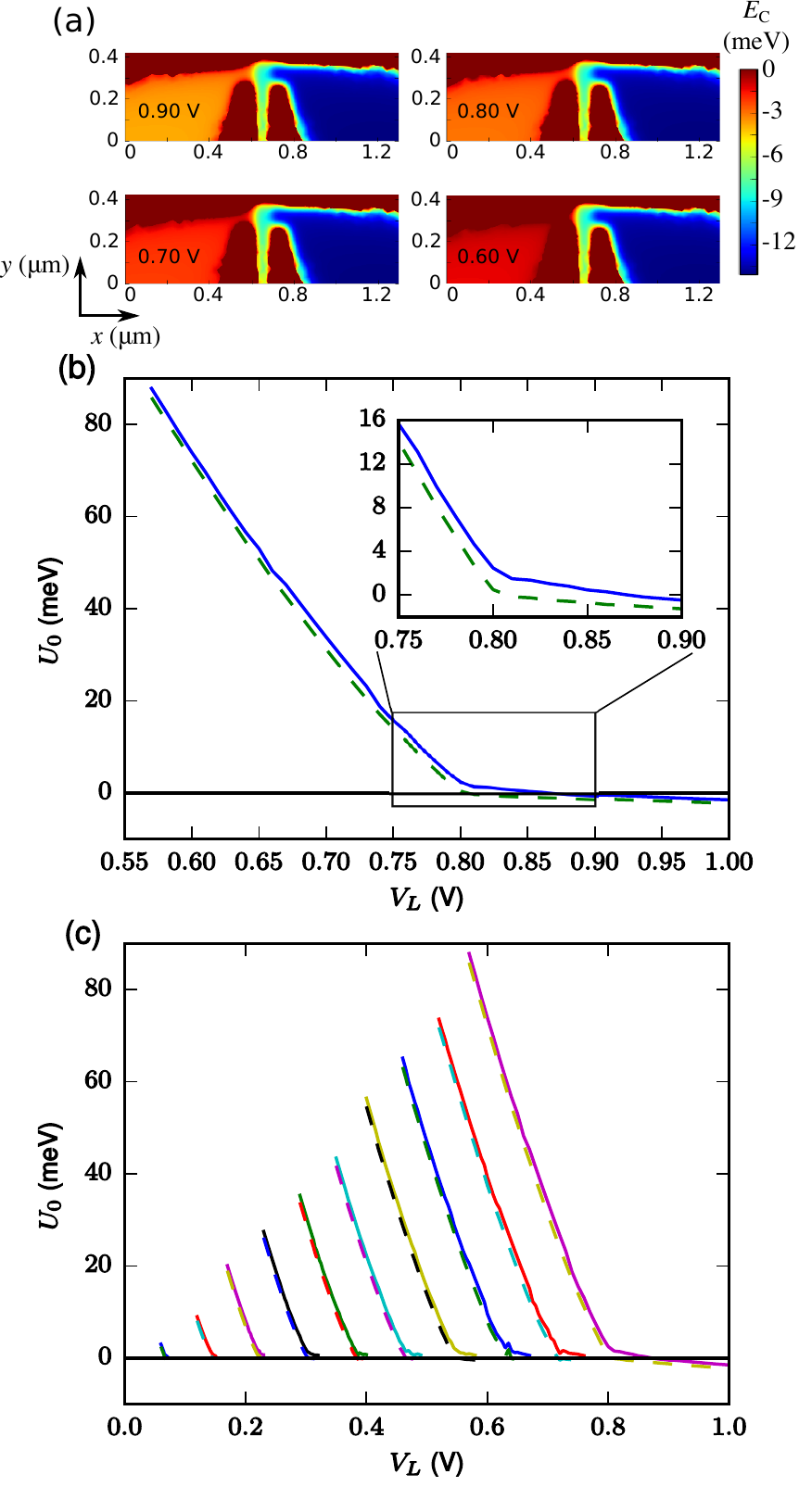} 
   \caption{ \label{fig4}3D electrostatics simulation of the tunnel barrier.
   (a): Conduction band energy plots for different values of $V_{L}$. The other gates are $V_{C} = 1.60$~V, and $V_{R} = 2.13$~V. The Fermi level is defined at zero in these plots. 
   (b): Simulated barrier height as a function of $V_{L}$ for $V_{C} = 1.60$~V, $V_{R} = 2.13$~V and an offset voltage of 0.50~V. Solid line shows barrier height values where the effect of quantum confinement has been accounted for while the dashed line does not take into account this effect. Inset: Zoom-in of boxed region to clarify effect of lateral confinement on barrier height. 
   (c): Barrier height as a function of $V_{L}$ for a range of offset voltages. The offset voltage varies from 0.05~V (leftmost curve) to 0.50~V (rightmost curve) in steps of 0.05~V. Solid and dashed lines have same meaning as in panel (b).}
\end{figure}

The width of the barrier is plotted as a function of $V_{L}$ for $V_{C} = 1.60$~V in Fig.~\ref{fig3}(d). In this case we show a two-parameter fit where the barrier height is not fixed. We see that the width appears to be a constant over the range of $V_{L}$ values considered. The average value is about 12~nm. We note that the extracted widths correspond well with our gate pattern.

In the context of compact models, the overall barrier height and width dependences on $V_{L}$ and $V_{C}$ have relatively simple forms that describe the behavior over very wide bias ranges. 

In addition to profile likelihood error bars, we also compute full 2D confidence regions for the parameters $U_{0}$ and $w$.
Similar to the 1D case, we assume that the difference between the MLE point and the likelihood function is distributed as $\chi^2(2)$.
Fig.~\ref{fig3}(e) shows the result for $V_{L}$ = 0.70~V, where the contours represent different confidence levels. 

Besides performing the parameter extractions above, we consider cases where the DOS in the leads is 0D, 1D, or 2D to examine if there are any distinguishing signatures in the calculated dependences. The difference in DOS affects the total number of available modes $M(E)$ in Eq.~\ref{Eq1}. All three cases fit the data well when the barrier heights and widths are allowed to adjust to compensate the change in the DOS.  
We note that the estimated widths of the leads would be consistent with transversal quantization and the energy level splitting is too large to warrant 1D DOS for the transmission modes.  We therefore use the 0D DOS for the extracted barrier heights and widths. More generally, it is unclear at this time what factors in the tunnel barrier geometry express clearer signatures due to the DOS dimensionality.  This is a topic for future examination.

We lastly compare the results of the 1D barrier model to electrostatic simulations of the electron density and electric potential (\emph{i.e.}, the conduction band edge, $E_{C}$). The electrostatics of the device is modeled using COMSOL Multiphysics software. We use the Thomas-Fermi approximation to model the electric potential and the electron density at the oxide-semiconductor interface with a 2D density of states for the experimental gate voltages. The electron density is calibrated at a gate voltage of 0.48~V by applying an offset voltage to all electrodes while ramping gate L. The offset voltage is used to ensure that the electron density in the leads agrees with the definition of threshold given by Eq.~\ref{density}.

Plots of the conduction band energy as a function of $V_{L}$ and position are shown in Fig.~\ref{fig4}(a). As $V_{L}$ is decreased a single barrier forms between gates L and C. The barrier height is determined by calculating the saddle point in the potential, which is found by calculating the gradient of the potential. We note that the shape of the barrier will generally not be trapezoidal for a 1D cut of the conduction band energy along a curve through the saddle point. The barrier height as a function of $V_{L}$ for $V_{C} = 1.60$~V, $V_{R} = 2.13$~V, and an offset voltage of 0.50~V is shown in Fig.~\ref{fig4}(b-c). 

In the constriction where the barrier forms we expect quantum confinement to have a significant effect. These transverse confinement energies were computed for this device geometry and become appreciable at the lower densities where the channel is narrow, see Appendix~\ref{sec:confinement}. The effect of lateral confinement is shown more clearly in the inset of Fig.~\ref{fig4}(b). This produces a region of relatively linear dependence of the barrier height on voltage after the electron channel is completely formed but the gap is not completely saturated with electron density.  A qualitatively similar region is seen in the extracted barrier heights in Fig.~\ref{fig3}(b), although extended over a wider voltage range. 
We note that better quantitative agreement is beyond the scope of this paper and probably requires self-consistent Schr\"{o}dinger-Poisson calculations.

Fig.~\ref{fig4}(c) shows the barrier height as a function of $V_{L}$ for a variety of offset voltages (\emph{i.e.}, threshold offsets for the COMSOL calculation). Each curve shows the range from no accumulation (large barrier) to a non-existent barrier (barrier height is zero or lower).

\section{Conclusion}
\label{sec:conclusion}
In conclusion, we have measured and modeled the transport spectroscopy of silicon MOS surface electrode-defined tunnel barriers at 4~K. Multiple electrodes are used to form the electrostatically defined barriers making for a large parameter space on which the tunnel barrier is dependent.  

We examine a 1D barrier model using voltage dependent barrier height and width, which is found to fit the current-voltage dependence well over a large parameter range.  The model is quasi-analytic providing both accurate 1D transmission values for the barrier model, while also being a fast extraction method to more readily enable its application to multi-electrode cases.  Cryogenic and quantum confinement effects are included in the model to account for threshold and barrier height shifts. 

The barrier height dependence on gate voltage was shown to be relatively linear in the high and low voltage regimes with an intermediate non-linear regime that is nearly exponential.  The regimes and their voltage dependences can be fit in a compact way offering a potential approach to modeling more complex nanostructures that have multiple tunnel barriers and more electrodes. 

We also compare the results of the 1D  model to semi-classical solutions of Poisson's equation using COMSOL and find that the simulations qualitatively agree well with the model predicted three regimes. 

Transport spectroscopy, itself, is a relatively rapid way to characterize tunnel barriers relative to more time intensive and complex pulsing approaches used for these kinds of barriers \cite{elzerman2004,amasha2008,thalakulam2011}. The combination of this method with transport spectroscopy offers a relatively rapid way to build a compact model of a tunnel barrier for multiple gate electrodes and reasonably large bias ranges.  This work should provide useful insight about the details of electrostatic barriers and how to characterize them. 

\begin{acknowledgments}
We gratefully recognize conversations with D.R.~Ward and P.A.~Sharma about this work and the manuscript. This work was performed, in part, at the Center for Integrated Nanotechnologies, a U.S. DOE, Office of Basic Energy Sciences user facility. The work was supported by the Sandia National Laboratories Directed Research and Development Program. Sandia National Laboratories is a multi-program laboratory operated by Sandia Corporation, a Lockheed-Martin Company, for the U. S. Department of Energy under Contract No. DE-AC04-94AL85000. 
\end{acknowledgments}

\begin{figure}
   \includegraphics{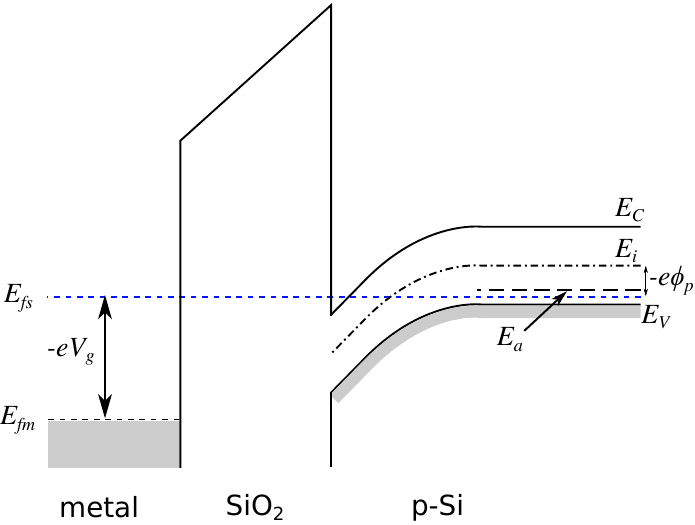} 
   \caption{ \label{energyDiagram} Low temperature energy band diagram for a MOS system with a p-type Si substrate biased into inversion. The quasi-Fermi level in the semiconductor $E_{fs}$ is pinned between the acceptor level $E_{a}$ and top of the valence band $E_{V}$ due to freeze-out.}
\end{figure}

\appendix
\section{Cryogenic threshold voltage calculation}
\label{sec:threshold}
We calculate the threshold voltage for a MOSFET at low temperatures by considering (1) the metal-insulator transition and (2) the consequences of low temperature on parameters used to calculate threshold, such as the metal-semiconductor work function, $\phi_{ms}$. First consider the metal-insulator transition.

One definition of threshold voltage for MOSFETs at high temperatures (such as room temperature) is that it is the gate voltage for which the mobile electron charge $Q_{n}=0$. As noted in the text, at cryogenic temperatures there is a metal-insulator transition where one must achieve a critical electron density $n_{crit}$ before appreciable conduction begins. Thus at low temperatures the threshold voltage is defined as the gate voltage for which $Q_{n}=-en_{crit}$. The expression for the threshold voltage is then given by 
\begin{equation}
V_{th}=V_{0}+V_{crit}. \label{threshold}
\end{equation}
In this expression $V_{0}$ is the standard MOSFET threshold voltage (discussed below) and $V_{crit}=en_{crit}/C_{ox}$, where $C_{ox}$ is the oxide capacitance per unit area. The critical density can be extracted from Hall measurements \cite{tracy2009}.

We next examine $V_{0}$. 
The usual formula for the threshold voltage, neglecting the body effect (that is, zero voltage difference between the source and bulk) and assuming a p-type substrate, is given by \cite{muller2002device}
\begin{equation}
V_{0}=V_{FB}+ 2\lvert\phi_{p}\rvert+\frac{1}{C_{ox}}\sqrt{4\epsilon_{s}eN_{a}\lvert\phi_{p}\rvert}, \label{threshold0}
\end{equation}
where $V_{FB}$ is the flat-band voltage, $\phi_{p}$ is the bulk potential (difference between the intrinsic and quasi-Fermi levels in the bulk of the semiconductor) and the third term is the voltage across the oxide due to the depletion layer charge. The flat-band voltage is given by $V_{FB}=\phi_{ms}-Q_{f}/C_{ox}$. The electrical permittivity of the semiconductor is $\epsilon_{s}$ and the acceptor concentration is $N_{a}$.

The parameters $\phi_{ms}$ and $\phi_{p}$ are most affected by changes in temperature. At cryogenic temperatures, the quasi-Fermi level in the semiconductor is pinned halfway between the acceptor energy levels and the top of the valence band due to freeze-out, as shown in Fig.~\ref{energyDiagram}. For boron, the acceptor energy levels are about 0.045~eV above $E_{V}$. It is well known that the band-gap energy $E_{g}$ changes with temperature; for Si, the band-gap energy is 1.17~eV at low temperatures. Thus we find $\phi_{p}=E_{g}/2 - 0.045/2=0.563$~V. 

The parameter $\phi_{p}$ also affects $\phi_{ms}$ through the semiconductor work function $\phi_{s}$. Unlike $E_{g}$, the effect of low temperatures on the metal work function $\phi_{m}$ and the electron affinity $\chi$ is not clear. We use $\phi_{m}=4.05$~V for n$^{+}$ poly-silicon and assume $\chi=4.05$~V for Si. This gives $\phi_{s}=\chi + E_{g}/2 + \phi_{p}=5.20$~V and a metal-semiconductor work function of $\phi_{ms}=\phi_{m}-\phi_{s}=-1.15$~V.

\section{Lateral quantum confinement}
\label{sec:confinement}
\begin{figure}
   \includegraphics{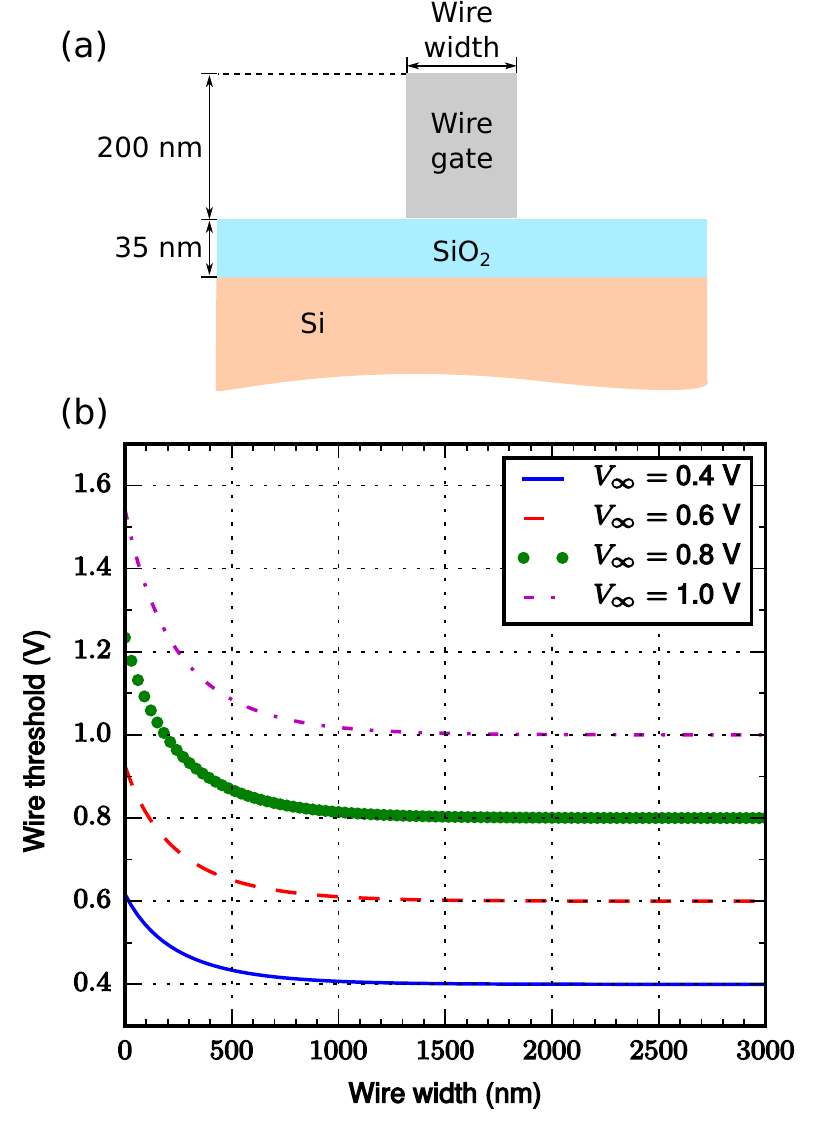} 
   \caption{ \label{thresholdShiftFig}(a) Cross-sectional schematic of the MOS geometry used to model effect of quantum confinement on threshold voltage. (b) Wire threshold voltage as a function of wire width for different field threshold voltages.}
\end{figure}

In a simple square potential well or particle in a box, the discrete energy levels increase as the width of the quantum well is reduced. Similarly, quantum confinement in a nanowire causes the lowest sub-band to increase in energy. In this appendix, we consider the effect of lateral quantum confinement on the threshold voltage. This effect causes a shift in the threshold voltage resulting in the threshold of a thin wire gate to be larger than the threshold of a much wider gate (field threshold). This theory was recently developed for nanowires in Si/SiGe heterostructures \cite{lu2016}; here, we apply it to the MOS geometry shown in Fig.~\ref{thresholdShiftFig}(a).

To understand this effect, we consider a range of wire gate voltages and wire widths. For each wire width, we compute the electrostatic potential for each corresponding wire gate voltage by solving Poisson's equation in 2D with the finite element method in COMSOL Multiphysics. We then calculate the ground state energy of the resulting confinement potential by solving Schr\"odinger's equation. This gives us the energy of the lowest sub-band as a function of voltage. We find that the ground state energy is linear with gate voltage, that is,
\begin{equation}
E_{0}^{L}=m_{L}V_{G}, \label{groundStateEnergy}
\end{equation}
where $E_{0}^{L}$ is the ground state energy of a wire of width $L$, $m_{L}$ is the slope or lever arm, and $V_{G}$ is the wire gate voltage. As the wire width increases, the slopes $m_{L}$ saturate. We then define the lever arm for an infinitely wide wire as approximately equal to the lever arm for a 5000~nm wide wire, that is, $m_{\infty}\approx m_{5000}=$ -0.91~meV/mV for the MOS geometry shown in Fig.~\ref{thresholdShiftFig}(a).

Assuming that conduction occurs when the ground state energy is aligned with some external reference energy level, $E^{0}$ (for example, the quasi-Fermi level), we define the field threshold voltage as,
\begin{equation}
V_{\infty}=\frac{E^{0}}{m_{\infty}}, \label{fieldThreshold}
\end{equation}
where $V_{\infty}$ is the threshold for an infinitely wide wire (field threshold).

If the threshold voltage of a wire of width $L$ is $V_{L}$ and we define the threshold shift as $\Delta V_{L}= V_{L}-V_{\infty}$, then we may write
\begin{eqnarray}
\Delta V_{L} & = & V_{L}-V_{\infty} \nonumber
\\
& = & \frac{E^{0}}{m_{L}} - \frac{E^{0}}{m_{\infty}} \nonumber
\\
& = & \frac{E^{0}m_{\infty}}{m_{L}m_{\infty}} - \frac{E^{0}m_{L}}{m_{\infty}m_{L}} \nonumber
\\
& = & \frac{E^{0}}{m_{L}m_{\infty}}\left(m_{\infty}-m_{L}\right). \label{thresholdShiftEq}
\end{eqnarray}
We can re-write the expression for the threshold shift explicitly in terms of the field threshold using Eq.~\ref{fieldThreshold} as,
\begin{equation}
\Delta V_{L}=V_{\infty}\left(\frac{m_{\infty}}{m_{L}}-1\right). \label{thresholdShiftEq2}
\end{equation}

Moreover, using the definition $\Delta V_{L}= V_{L}-V_{\infty}$, we can write a direct expression for the wire threshold as,
\begin{equation}
V_{L}=V_{\infty}\frac{m_{\infty}}{m_{L}}. \label{wireThreshold}
\end{equation}
We find that the lever arm $m_{L}$ as a function of wire width $L$ is fit well by a decaying exponential function, that is, $m_{L}\approx m_{\infty} + ae^{-bL}$, where $a=0.32$ meV/mV and $b=0.003$ nm$^{-1}$. The expression for the wire threshold Eq.~\ref{wireThreshold} then becomes,
\begin{equation}
V_{L}=V_{\infty}\frac{m_{\infty}}{m_{\infty} + ae^{-bL}}. \label{wireThreshold2}
\end{equation}

Thus we can calculate the wire threshold given a field threshold voltage and wire width using this compact model. The wire threshold as a function of wire width for different field thresholds is shown in Fig.~\ref{thresholdShiftFig}(b). We note this assumes that lateral quantum confinement is the dominant effect and the exact increase due to confinement is geometry dependent.

\section{Transport formalism for a general trapezoidal barrier}
\label{sec:formalism}
In this appendix, we present a detailed calculation of tunneling through a ``1D" trapezoidal barrier. We will use the Landauer-B\"uttiker formalism, computing the transmission function via a semi-analytical approach.

The Landauer-B\"uttiker formalism is a general framework for computing tunneling quantum transport through a device. It can handle devices with many leads and high source-drain biases. The method breaks down when transport becomes non-ballistic, through ``vertical" scattering events. This occurs when energy is not conserved throughout the scattering region; in this instance, more sophisticated techniques, such as non-equilibrium Green's functions, must be employed. 

The case we consider here is a simple, two-terminal device. Within the Landauer-B\"uttiker formalism, the total current through the scattering region is given by
\begin{equation}
I = \int dE \, \frac{2 e}{h} \bar T(E) \left( f_{s}(E) - f_{d}(E) \right), \label{totalCurrent}
\end{equation}
where $\bar T(E)$ is the total transmission function at a given energy (summed up over transverse modes), $f_s$ is the Fermi function of the source, and $f_d$ is the Fermi function of the drain. The difficult part of computing $I$ is to calculate the total transmission function $\bar T(E)$. To facilitate this, we decompose $\bar T$ as
\begin{align}
\bar T(E) &= \sum_{m \in modes} T_m(E) \nonumber \\
& = T(E)  M(E),
\end{align}
where $T_m(E)$ is the transmission of a given transverse mode, $M(E)$ is the total number of accessible modes at a given energy $E$. Hence, we have decomposed the problem into computing $T(E)$ and $M(E)$ indpendently. First, we examine the computation of $T(E)$.

\subsection{Computing the transmission function}
\begin{figure}[tb]
\includegraphics[width= 0.95 \linewidth]{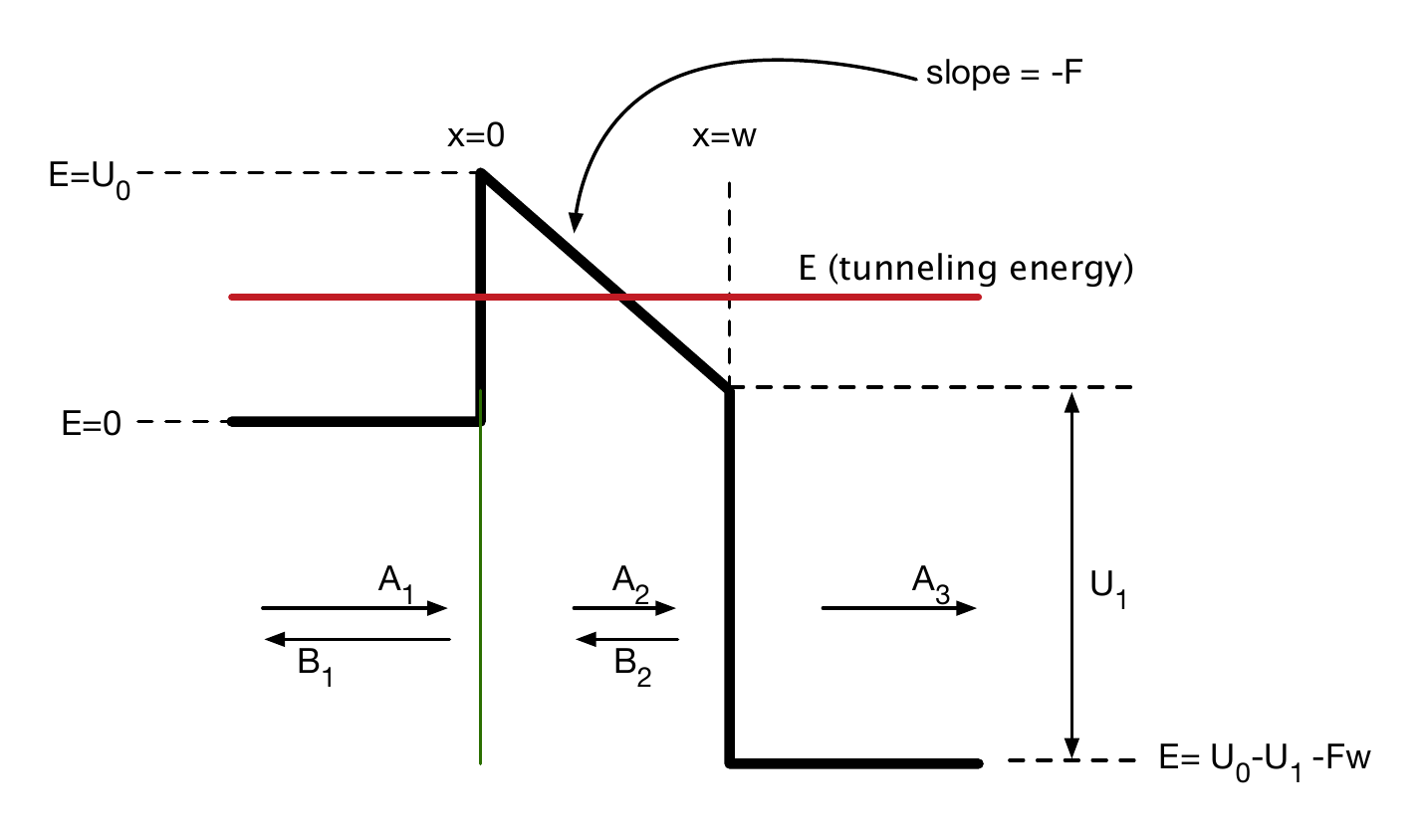}
\caption{\label{transmission}Schematic of the transmission problem.
}
\end{figure}

The transmission function gives the probability of an incident mode tunneling through the barrier. Schematically, we consider the case shown in Fig.~\ref{transmission}. There, the transmission function is a function of the energy $E$ and is parameterized by the left-side barrier height $U_0$, the field $F$ between the source and drain, the barrier width $w$, and the right-side barrier height $U_1$. 

\subsubsection{Form of the Schr\"odinger equation}

To compute the transmission function, we need to solve Schr\"odinger's equation for an incoming plane wave at a specified energy. Within each region, we can write down the solution analytically:
\begin{align}
\psi_1(x) &= A_1 e^{i k_1 x} + B_1 e^{-i k_1 x},\\
\psi_2(x) &= A_2 \text{Ai}\left(a(x)\right)
            +B_2 \text{Bi}\left(a(x)\right),\\
\psi_3(x) &= A_3 e^{i k_3 x}.
\end{align}
Here, we have $k_1 = \frac{\sqrt{2 m E}}{\hbar}$, $k_3 = \frac{\sqrt{2 m (E-(U0-U1-Fw))}}{\hbar}$,
and $a(x) = \left[\frac{2 m F}{\hbar^2}\right]^{1/3} \left[\frac{U_0 - E}{F}-x \right]$.
The transmission function is given by $T = \left| \frac{A_3}{A_1} \right|^2$.
To solve for this quantity, we need to impose boundary conditions: $\psi_1(0) = \psi_2(0)$, $\psi_1'(0) = \psi_2'(0)$, $\psi_2(w) = \psi_3(w)$, $\psi_2'(w) = \psi_3'(w)$. To satisfy the boundary conditions, we will also need to know the formulas for the derivatives:
\begin{align}
\psi_1'(x) &= A_1i k_1 e^{i k_1 x} - B_1i k_1 e^{-i k_1 x},\\
\psi_2'(x) &= -\left(\frac{2 F m }{\hbar^2}\right)^{1/3} \left[A_2 \text{Ai}' \left(a(x)\right)
            +B_2 \text{Bi}'\left(a(x)\right)\right], \nonumber \\
\psi_3'(x) &= A_3 i k_3 e^{i k_3 x}. \nonumber 
\end{align}

\subsubsection{Renormalizing the Airy functions}

One issue that arises immediately is that Airy functions that form the solution in region 2 are not well-defined when $F\rightarrow 0$. This is a problem, since we want to recover the original square barrier problem in this case. Since the argument of the Airy functions is real, we can use the asymptotic form for $F\rightarrow 0$:
\begin{widetext}
\begin{align}
\mbox{Ai}(z) &\sim \lim_{N\rightarrow \infty} \mbox{Ai}_N(z) \equiv 
\frac{e^{-2/3 z^{3/2}}}{2 \sqrt{\pi} z^{1/4}}
\sum_{n=0}^N \frac{(-1)^n \Gamma\left(n + \frac{5}{6}\right) 
\Gamma \left(n+\frac{1}{6} \right) \left(\frac{3}{4} \right)^n}
{2 \pi n! z^{3 n/2}},\\
\mbox{Bi}(z) &\sim \lim_{N\rightarrow \infty} \mbox{Bi}_N(z) \equiv 
\frac{e^{2/3 z^{3/2}}}{ \sqrt{\pi} z^{1/4}}
\sum_{n=0}^N \frac{ \Gamma\left(n + \frac{5}{6}\right) 
\Gamma \left(n+\frac{1}{6} \right) \left(\frac{3}{4} \right)^n}
{2 \pi n! z^{3 n/2}}.
\end{align}
\end{widetext}
These expressions are valid for real, large-magnitude $z$. We want to use this to work out expressions for the boundary conditions of $\psi_2(x)$. We will use these to renormalize the solution coefficients so that we don't have undefined expressions. Numerically, we find that we need only go to $N=0$ to obtain good accuracy. 

We write down the asympotic result at $x=0$:
\begin{align}
\text{Ai}_{N=0}(a(0))&\equiv \alpha_0 = 
\frac{\left(F \hbar \right)^{1/6} e^{-\frac{2 \sqrt{2 m}
   (U_0-E )^{3/2}}{3 F \hbar }}}{2 
       \sqrt{\pi } \left( 2 m\right)^{1/12} \left(\text{U0}-E \right)^{1/4}},\\
\text{Bi}_{N=0}(a(0))&\equiv \beta_0 = 
\frac{\left(F \hbar \right)^{1/6} e^{\frac{2 \sqrt{2 m}
   (U_0-E )^{3/2}}{3 F \hbar }}}{ 
       \sqrt{\pi } \left( 2 m\right)^{1/12} \left(\text{U0}-E \right)^{1/4}}.
\end{align}
Using this, we rewrite the $\psi_2$ solution as:
\begin{align}
\psi_2(x) &= \left(A_2 \alpha_0 \right) \frac{\text{Ai}\left(a(x)\right)}{\alpha_0} \nonumber
 +\left(B_2 \beta_0\right) \frac{\text{Bi}\left(a(x)\right)}{\beta_0} \\
 &\equiv \mathcal{A}_2 \mathcal{A}i\left(a(x)\right) + 
 \mathcal{B}_2\mathcal{B}i\left(a(x)\right),
\end{align}
where the new functions $\mathcal{A}i$ and $\mathcal{B}i$ have the property that 
\begin{align}
\lim_{F\rightarrow 0}\mathcal{A}i\left(a(x)\right) = e^{\frac{x}{\hbar}\sqrt{2 m (U_0 - E)}},\\
\lim_{F\rightarrow 0}\mathcal{B}i\left(a(x)\right) = e^{-\frac{x}{\hbar}\sqrt{2 m (U_0 - E)}},
\end{align}
which are of course just the growing and decaying exponentials that span the solution for a flat barrier.

Hence, this renormalization allows us to make contact with the simple limiting case. We thus replace our region 2 wave function and derivative with following piecewise definitions:
\begin{widetext}
\begin{align}
\psi_2(x) &=
\begin{cases}
A_2 \frac{\text{Ai}\left(a(x)\right)}{\alpha_0} + B_2 \frac{\text{Bi}\left(a(x)\right)}{\beta_0}  & F\geq F_0 \\
A_2 e^{\frac{x}{\hbar}\sqrt{2 m (U_0 - E)}} + B_2 e^{-\frac{x}{\hbar}\sqrt{2 m (U_0 - E)}}  & F<F_0
\end{cases},\\
\psi_2'(x) &=
\begin{cases}
-\left(\frac{2 F m }{\hbar^2}\right)^{1/3} \left[A_2 \frac{\text{Ai}' \left(a(x)\right)}{\alpha_0}
            +B_2 \frac{\text{Bi}'\left(a(x)\right)}{\beta_0}\right]  & F\geq F_0 \\
\frac{\sqrt{2 m (U_0 - E)}}{\hbar} \left[A_2 e^{\frac{x}{\hbar}\sqrt{2 m (U_0 - E)}} - B_2 e^{-\frac{x}{\hbar}\sqrt{2 m (U_0 - E)}}\right]  & F<F_0
\end{cases}.
\end{align}
\end{widetext}
Here, we just relabeled the unknown coefficients $A_2$ and $B_2$ that we are solving for to include the renormalization. We identify a good value of a threshold field $F_0$ numerically.

\subsubsection{Constructing the system of equations}

Now that we have worked out the wave functions for all three regions, we are ready to compute the transmission function. However, notice that we have five unknown coefficients with only four constraints. This is because the full wave function needs to be normalized, which provides the fifth constraint. However, the transmission function only needs a ratio of coefficients, so we don't need to worry about this additional constraint. Specifically, we divide through everywhere by $A_1$, giving:
\begin{widetext}
\begin{align}
\psi_1(x) &= e^{i k_1 x} + \tilde{B_1} e^{-i k_1 x},\\
\psi_2(x) &=
\begin{cases}
\tilde{A_2}\frac{\text{Ai}\left(a(x)\right)}{\alpha_0} + \tilde{B_2} \frac{\text{Bi}\left(a(x)\right)}{\beta_0}  & F\geq F_0 \\
\tilde{A_2} e^{\frac{x}{\hbar}\sqrt{2 m (U_0 - E)}} + \tilde{B_2} e^{-\frac{x}{\hbar}\sqrt{2 m (U_0 - E)}}  & F<F_0
\end{cases},\\
\psi_3(x) &= \tilde{A_3} e^{i k_3 x},\\
\psi_1'(x) &= i k_1 e^{i k_1 x} - \tilde{B_1}i k_1 e^{-i k_1 x},\\
\psi_2'(x) &=
\begin{cases}
-\left(\frac{2 F m }{\hbar^2}\right)^{1/3} \left[\tilde{A_2} \frac{\text{Ai}' \left(a(x)\right)}{\alpha_0}
            +\tilde{B_2} \frac{\text{Bi}'\left(a(x)\right)}{\beta_0}\right]  & F\geq F_0 \\
\frac{\sqrt{2 m (U_0 - E)}}{\hbar} \left[\tilde{A_2} e^{\frac{x}{\hbar}\sqrt{2 m (U_0 - E)}} - \tilde{B_2} e^{-\frac{x}{\hbar}\sqrt{2 m (U_0 - E)}}\right]  & F<F_0
\end{cases},\\
\psi_3'(x) &= \tilde{A_3} i k_3 e^{i k_3 x}.
\end{align}
\end{widetext}
Here, the tilde variables indicate scaling by $A_1$, so $T = \left|\tilde{A_3} \right|^2 k_3/k_1$, where the last factor comes from the difference in velocity between incoming and outgoing modes. We will now drop the tildes for convenience.

Considering first the case where the Airy functions are well-defined, we have the following system of equations to solve:
\begin{widetext}
\begin{align}
1 &= -1 B_1 + \frac{\text{Ai}\left(a(0)\right)}{\alpha_0}  A_2 + \frac{\text{Bi}\left(a(0)\right)}{\beta_0} B_2 + 0 A_3,\\
i k_1 &=  i k_1 B_1 
          -\left[ \left(\frac{2 F m }{\hbar^2}\right)^{1/3}\frac{\text{Ai}' \left(a(0)\right)}{\alpha_0}\right] A_2
          -\left[ \left(\frac{2 F m }{\hbar^2}\right)^{1/3}\frac{\text{Bi}' \left(a(0)\right)}{\beta_0}\right] B_2
          + 0 A_3, \nonumber \\
0 &= 0 B_1 +\left[ \frac{\text{Ai}\left(a(w)\right)}{\alpha_0} \right] A_2
      +\left[ \frac{\text{Bi}\left(a(w)\right)}{\beta_0} \right] B_2
      - \left[e^{i k_3 w}\right] A_3,\nonumber \\
0 &= 0 B_1 
        +\left[ \left(\frac{2 F m }{\hbar^2}\right)^{1/3}\frac{\text{Ai}' \left(a(w)\right)}{\alpha_0}\right] A_2
          +\left[ \left(\frac{2 F m }{\hbar^2}\right)^{1/3}\frac{\text{Bi}' \left(a(w)\right)}{\beta_0}\right] B_2
      + \left[i k_3 e^{i k_3 w}\right] A_3.\nonumber      
\end{align}
If instead the Airy functions are not well defined and we have to use the asymptotic forms, we have:
\begin{align}
1 &= -1 B_1 + 1  A_2 + 1 B_2 + 0 A_3,\\
i k_1 &=  i k_1 B_1 
          +\left[ \frac{\sqrt{2 m (U_0 - E)}}{\hbar}\right] A_2
          -\left[ \frac{\sqrt{2 m (U_0 - E)}}{\hbar}\right] B_2
          + 0 A_3, \nonumber \\
0 &= 0 B_1 +\left[ e^{\frac{w}{\hbar}\sqrt{2 m (U_0 - E)}} \right] A_2
      +\left[ e^{-\frac{w}{\hbar}\sqrt{2 m (U_0 - E)}} \right] B_2
      - \left[e^{i k_3 w}\right] A_3,\nonumber \\
0 &= 0 B_1 
        -\left[ \frac{\sqrt{2 m (U_0 - E)}}{\hbar}e^{\frac{w}{\hbar}\sqrt{2 m (U_0 - E)}} \right] A_2
          +\left[  \frac{\sqrt{2 m (U_0 - E)}}{\hbar}e^{-\frac{w}{\hbar}\sqrt{2 m (U_0 - E)}}\right] B_2
      + \left[i k_3 e^{i k_3 w}\right] A_3.\nonumber      
\end{align}
\end{widetext}
We cast these equations as a simple matrix problem solve for $A_3$.

\subsection{Computing the total current}
\begin{figure}[tb]
\includegraphics[width= 0.95 \linewidth]{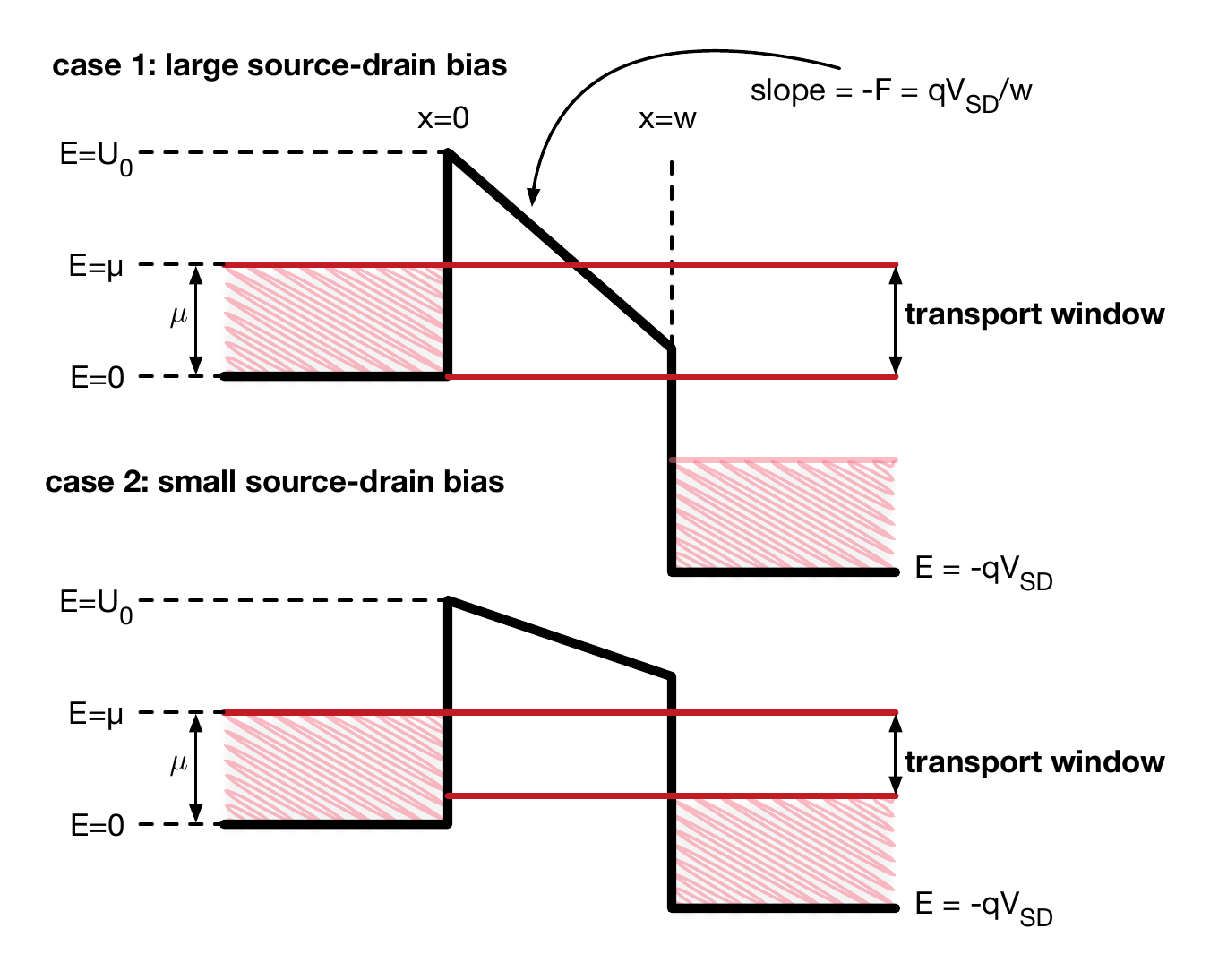}
\caption{\label{transport}Schematic for computing the total current.
}
\end{figure}

Now that we can reliably compute the transmission function, we need to use it to compute the total current within the Landauer-B\"uttiker formalism. To do this, we modify the schematic slightly to Fig.~\ref{transport}.

\subsubsection{Density of transverse modes}

The total transmission function $\bar T(E)$ is given by 
\begin{align}
\bar T(E) &= \sum_{m \in modes} T_m(E) \\
& = T(E)  M(E) \nonumber \\
& = T(E) L^d\int_0^{E}dE' g_d(E'), \nonumber
\end{align}
where $T_m(E)$ is the transmission of a given transverse mode, $M(E)$ is the total number of accessible modes at a given energy $E$, $g$ is the density of states, $L$ is the cross-sectional length scale of the scattering region, and $d$ is the dimension of the lead. The three cases of interest are for 0D (one mode), 1D (sheet contacts), and 2D (volume contacts) density of states, given by:
\begin{align}
g_0(E) &= \delta(E_0-E), \\
g_1(E) &= \frac{2}{\pi \hbar} \sqrt{\frac{2 m}{E}}, \nonumber\\
g_2(E) &= \frac{2 m}{\pi \hbar^2},\nonumber
\end{align}
where we have multiplied the usual formulas for the 1D and 2D DOS expressions by 2 for the extra valley degeneracy in silicon. Using these density of states formulas, we compute $M(E)$:
\begin{align}
M_0(E) & = 
\begin{cases} 
1 & E_0\leq E \\
0 & E_0 > E 
\end{cases} ,\nonumber\\
M_1(E) &= \frac{4 L \sqrt{2 m E}}{\pi \hbar}, \\
M_2(E) &= \frac{2 m L^2 E}{\pi \hbar^2}.\nonumber
\end{align}

\subsubsection{Computing the current}
Equipped with the number of modes, we can compute current:
\begin{align}
I &= \int dE \, \frac{2 e}{h} \bar T(E) \left( f_{s}(E) - f_{d}(E) \right) \\
  & \approx \frac{2 e}{h} \int_{\mu_0}^{\mu} dE T(E) M(E), \nonumber 
\end{align}
where
\begin{equation}
\mu_0 = \begin{cases}
0 & \mu- q V_\mathrm{SD}<0\\
\mu- q V_\mathrm{SD} & \mu- q V_\mathrm{SD} \geq 0
\end{cases}.
\end{equation}
The two cases here correspond to the case 1 and case 2 in Fig.~\ref{transport}. From the formulas above, we can write three different current models:
\begin{align}
I_0 &= \frac{2 e}{h} \int_{\mu_0}^{\mu} dE T(E), \\
I_1 &= \frac{2 e}{h} \int_{\mu_0}^{\mu} dE \frac{4 L  T(E)\sqrt{2 m E}}{\pi \hbar},\\
I_2 &= \frac{2 e}{h} \int_{\mu_0}^{\mu} dE \frac{2 m L^2 E T(E)}{\pi \hbar^2},
\end{align}
where we assume that for $I_0$ the single mode is in the transport window.

\begin{figure}
   \includegraphics{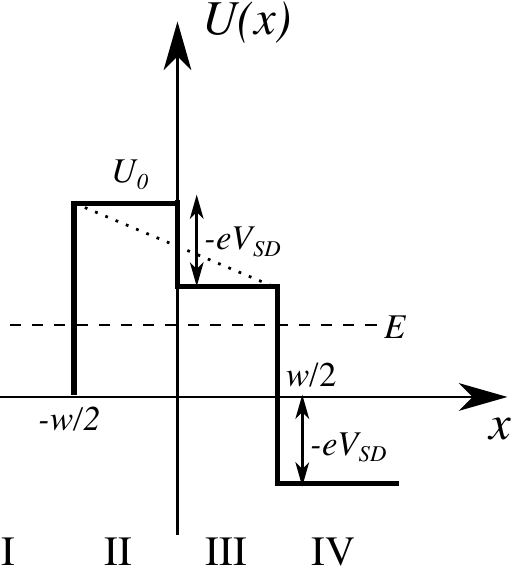} 
   \caption{ \label{threeStep} 1D potential energy $U(x)$ diagram showing three step barrier model. The energy of the impinging electron is $E$.}
\end{figure}

\section{Three-step model}
\label{sec:threeStep}
In the three-step barrier model we approximate a square barrier under bias using three step potentials as shown in Fig.~\ref{threeStep}. For $V_{SD}=0$, the barrier reduces to a square barrier of height $U_{0}$ and width $w$. The first step occurs at $x=-w/2$ and has a fixed height of $U_{0}$. The next step occurs at $x=0$ and has a variable step height that depends on the source-drain bias $V_{SD}$. The final step occurs at $x = w/2$ and also has a variable height that depends on $V_{SD}$. We use the transfer or $T$-matrix formalism to compute the transmission coefficient \cite{davies1997,ando1987}. For simplicity we only consider positive $V_{SD}$ but this approach can be easily applied to negative $V_{SD}$ as well.

First consider the case where $E>U_{0}$. There are four wavenumbers, each corresponding to a particular region of the barrier as indicated by the roman numerals in Fig.~\ref{threeStep}. The wavenumbers are given as follows: 
\begin{eqnarray}
k_{1}^{2}&=&2mE/\hbar^{2}, \label{k1}
\\ 
k_{2}^{2}&=&2m\left(E-U_{0}\right)/\hbar^{2}, \label{k2}
\\
k_{3}^{2}&=&2m\left(E-U_{0}+eV_{SD}\right)/\hbar^{2}, \label{k3} 
\\
k_{4}^{2}&=&2m\left(E+eV_{SD}\right)/\hbar^{2}. \label{k4}
\end{eqnarray}
Here $m$ is the effective mass of the electron. Using the $T$-matrix formalism we find the transmission coefficient to be:
\begin{widetext}
\begin{equation}
T= \frac{4k_{4}/k_{1}}{\left[f_{1}\cos\theta\cos\phi-f_{2}\sin\theta\sin\phi\right]^{2}+\left[f_{3}\sin\theta\cos\phi+f_{4}\cos\theta\sin\phi\right]^{2}}, \label{TFunc3Step}
\end{equation}
\end{widetext}
where the coefficients $f_{1}$, $f_{2}$, $f_{3}$, and $f_{4}$ are functions of the wavenumbers given by 
\begin{align}
f_{1}&=1+\frac{k_{4}}{k_{1}}, \label{f1}
\\
f_{2}&=\frac{k_{3}}{k_{2}}+\frac{k_{2}k_{4}}{k_{1}k_{3}}, \label{f2}
\displaybreak \\
 f_{3}&=\frac{k_{4}}{k_{2}}+\frac{k_{2}}{k_{1}}, \label{f3}
 \\
 f_{4}&=\frac{k_{4}}{k_{3}}+\frac{k_{3}}{k_{1}}. \label{f4}
\end{align}
The arguments of the trigonometric functions depend on the width of the barrier as $\theta = k_{2}w/2$ and $\phi = k_{3}w/2$. This is the general form of the transmission coefficient.

Now consider the large $V_{SD}$ case such that $V_{SD}>U_{0}-E$ and $U_{0}>E$. The form of the transmission coefficient and the wavenumbers remain the same except the wavenumber in region II is replaced in the following manner: $k_{2} \rightarrow i\kappa_{2}$. Then the trigonometric functions that have $\theta$ as the argument become hyperbolic functions, that is $\sin k_{2}w/2 \rightarrow i\sinh \kappa_{2} w/2$ and $\cos k_{2}w/2 \rightarrow \cosh \kappa_{2} w/2$. The wavenumber in region II is now given by $\kappa_{2}^{2}=2m\left(U_{0}-E\right)/\hbar^{2}$.  

Then consider the small $V_{SD}$ case such that $V_{SD}<U_{0}-E$ and $U_{0}>E$, as shown in Fig.~\ref{threeStep}. Again the form of the transmission coefficient remains the same except now the wavenumbers in both region II and III are replaced: $k_{2} \rightarrow i\kappa_{2}$ and $k_{3} \rightarrow i\kappa_{3}$. The wavenumbers in region II and III are now given by $\kappa_{2}^{2}=2m\left(U_{0}-E\right)/\hbar^{2}$ and $\kappa_{3}^{2}=2m\left(U_{0}-E-eV_{SD}\right)/\hbar^{2}$ . Now all the trigonometric functions become hyperbolic functions as discussed above.

Finally Eq.~\ref{TFunc3Step} and the appropriate wavenumbers (based on the values of $V_{SD}$, $U_{0}$, and $E$) are used in Eq.~\ref{totalCurrent} to compute the current through the device.

\end{document}